\documentclass[12pt,tightenlines,eqsecnum,floats,showpacs,nofootinbib,amsmath,amssymb,aps,prd,
 superscriptaddress]{revtex4-1}

\usepackage{graphicx}
\usepackage{amsmath,amssymb}
\usepackage{hyperref}

\begin{document}

\title{Matter in inhomogeneous loop quantum cosmology: the Gowdy $T^3$ model}

\author{Mercedes Mart\'in-Benito}
\email{mercedes@aei.mpg.de}
\affiliation{MPI f\"ur Gravitational Physics, Albert Einstein Institute,
Am M\"uhlenberg 1, D-14476 Potsdam, Germany}
\affiliation{Instituto de Estructura de la Materia, CSIC,
Serrano 121, 28006 Madrid, Spain}
\author{Daniel \surname{Mart\'in-de~Blas}}
\email{daniel.martin@iem.cfmac.csic.es}
\affiliation{Instituto de Estructura de la Materia, CSIC,
Serrano 121, 28006 Madrid, Spain}
\author{Guillermo A. Mena Marug\'an} \email{mena@iem.cfmac.csic.es}
\affiliation{Instituto de Estructura de la Materia, CSIC,
Serrano 121, 28006 Madrid, Spain}

\begin{abstract}
We apply a hybrid approach which combines loop and Fock quantizations to fully
quantize the linearly polarized Gowdy $T^3$ model in the presence of a
massless scalar field with the same symmetries as the metric. Like in the absence
of matter content, the application of loop
techniques leads to a quantum resolution of the classical cosmological
singularity. Most importantly, thanks to the inclusion of matter, the homogeneous sector
of the model contains flat Friedmann-Robertson-Walker (FRW) solutions,
which are not allowed in vacuo. Therefore, this model provides a simple setting to
study at the quantum level interesting physical phenomena such as the effect
of the anisotropies and inhomogeneities on flat FRW cosmologies.
\end{abstract}


\maketitle

\section{Introduction}
\label{s1}

Loop quantum cosmology (LQC) \cite{lqc1,lqc2,lqc3,abl}
is a quantization of cosmological models inspired in loop
quantum gravity ideas and methods \cite{lqg1,lqg3}, in which the geometry
has a discrete quantum nature. The first model successfully quantized to
completion in LQC was the
flat FRW model minimally coupled to a massless scalar field, whose dynamical
analysis shows that a quantum bounce replaces the initial singularity \cite{aps}. The
resolution of the cosmological singularity is a robust property of the theory
\cite{acs,mmo}, owing to the polymeric representation adopted for the geometry,
and it is also achieved in the rest of homogeneous models quantized so far in LQC (see for
instance \cite{apsv,skl,kv,bp,chiou,mbmmp,awe1,awe2,we} and references
therein).

In order to allow for the presence of inhomogeneities within the
framework of LQC, recently a hybrid approach to the quantization has been
developed  in the example of the simpler case: the Gowdy
$T^3$ model with linear polarization \cite{hybrid1,hybrid2,hybrid3,hybrid4}.
This is a midisuperspace with three-torus spatial topology that contains
inhomogeneities varying in a single direction~\cite{gowdy}.

The introduced hybrid approach combines the techniques of LQC
with those of the Fock quantization for reduced models in which only
global constraints remain to be imposed at the quantum level. The phase
space is split in homogeneous and inhomogeneous sectors. The former is
described by the degrees of freedom that parameterize the subset of
homogeneous solutions, and the second one is formed by the rest of
degrees of freedom. In the quantum theory, the
inhomogeneous sector is represented \`a la Fock, in order to deal with the field
complexity, while the homogeneous sector is represented following LQC, with the
aim at obtaining a quantum model with no analog of the classical cosmological
singularity. The approach assumes a hierarchy of quantum phenomena, so that the
most relevant effects of the loop quantum geometry are those that affect the
homogeneous degrees of freedom. In the case of the quantized Gowdy model, the
homogeneous sector coincides with the phase space of the Bianchi I model, which
has been extensively studied in LQC \cite{bojo,chiou,mbmmp,awe1}.
Concerning
the inhomogeneous sector, the requirement that the conventional description
for the inhomogeneities should be recovered when the quantum geometry effects of
the homogeneous sector are negligible and that this description respect unitarity
selects the Fock quantization of Refs. \cite{ccm1,ccm2,ccmv2} without ambiguity.
In fact, with the commented requirement, it has been shown that this is
the unique satisfactory Fock quantization that the totally deparameterized
Gowdy $T^3$ model admits \cite{ccmv1,cmv}.

Our aim is to further analyze inhomogeneous cosmologies in LQC by means of this
hybrid quantization, now allowing for the presence of matter. In order to do
this, we will include in the Gowdy $T^3$ model a minimally coupled massless
scalar field with the same symmetries of the geometry. Choosing suitable field
parameterizations for the inhomogeneities of both gravitational
waves and matter, the corresponding field contributions appear in the constraints
in the same way \cite{bgv,bgvv}. As a consequence, the uniqueness results of Refs.
\cite{ccmv1,cmv} for the Fock quantization apply to the nonvacuum case
as well, and hence we have at our disposal a preferred Fock
description also for the inhomogeneities of the matter field.

The interest of this work lies not only in the fact that it provides a
complete quantization of a cosmological model with an inhomogeneous matter field
in the framework of LQC, but also in that it means a further step towards
the quantum analysis of physical inhomogeneities in cosmology, in the sense that
these inhomogeneities propagate on a geometry not very different from that of our
universe. Indeed, thanks to the inclusion of matter, now the homogeneous sector
of the model (nonvacuum Bianchi I) admits flat FRW cosmologies as a subset
of solutions, namely the isotropic ones, and it is widely known that the observed
universe can be approximated at large scales by a spacetime of this type. Therefore,
it is natural to compare the dynamics of our
inhomogeneous model with that of the flat isotropic model, and analyze the
quantum effects that anisotropies and inhomogeneities produce over a
hypothetical FRW-like background. In particular, this setting would allow us to
investigate questions like the robustness of the quantum bounce scenario of
LQC when inhomogeneities are included, or modifications to
the evolution of the matter inhomogeneities when quantum geometry effects
are taken into account.

Let us mention that, owing to the isometries of the Gowdy model, this family
of spacetimes presents a particular subset of solutions with local rotational
symmetry (LRS), in which the two scale
factors of the directions of homogeneity coincide. Therefore, in
order to simplify the analysis, it is convenient to focus on this kind of
solutions, which we will call LRS-Gowdy cosmologies in the following.
We will carry out this LRS reduction at the quantum level using an adaptation
of the (so-called) {\sl projection} procedure introduced in Ref.~\cite{awe1} to pass
from the loop quantized Bianchi I model to the loop quantized FRW model.

This work is intended as a first contribution to the analysis of the Gowdy system with
matter.
Specifically, here we quantize to completion the model, putting special attention
to the new features that the consideration of the matter field introduces in
comparison with the vacuum case. We also present a general discussion of the lines
of attack that can be pursued to extract the physics from our quantum model.
We leave for a future work a more rigorous and deeper study of this underlying
physics and its consequences. The structure of this paper is as follows.
The classical model is described in Sec. \ref{s2}. In Sec. \ref{s3}, we carry
out its quantization, promoting the constraints to operators and characterizing
the physical Hilbert space. We also show there how to reduce the quantum model
to the corresponding LRS-Gowdy counterpart. Finally, in the concluding
Sec. \ref{s4}, we point out the possibilities that this model provides to
analyze quantum phenomena in cosmology and to reach physical predictions.

\section{Classical model}
\label{s2}

The linearly polarized Gowdy $T^{3}$ cosmologies are globally hyperbolic
spacetimes with three-torus spatial topology and two axial and hypersurface
orthogonal Killing vector fields \cite{gowdy}. We provide this system with
a minimally coupled massless scalar field, $\Phi$, with the same symmetries.
We use global coordinates $\lbrace t, \theta, \sigma, \delta \rbrace$, where
$\theta,\sigma, \delta \in S^{1}$, such that the Killing fields are
$\partial_{\sigma}, \partial_{\delta}$. Then, all the fields (metric and matter)
only depend on the coordinates $t$ and $\theta$. We reduce the system by
performing a partial gauge fixing, as in Refs. \cite{ccm2,hybrid2,hybrid3}.
As a result, the gravitational sector of the phase space turns out to be described
by two pairs of canonically conjugate point-particle variables (they do not depend
on $\theta$) and by one field, together with its canonical momentum. We expand
the fieldlike variables in Fourier series in the coordinate $\theta$ and split
the phase space into two sectors: one formed by all the homogeneous degrees of
freedom (the two point-particle gravitational variables and the zero modes of
both matter and gravitational fields, together with their momenta) and the other
formed by the nonzero modes of the two fields of the system and of their conjugate
momenta. We call them homogeneous and inhomogeneous sectors, respectively.

In the totally deparameterized model, there is a particular field parameterization
of the metric in which the gravitational wave content is described by a
field which behaves exactly as the matter field $\Phi$, namely, as a massless scalar
field propagating in 2+1 gravity \cite{pierri}. Nonetheless, this description does
not admit any Fock quantization with unitary dynamics \cite{ccq,cm,cmv}. For that,
it is necessary to apply a time dependent canonical transformation on both
fields \cite{ccm2,cmv}. The resulting gravitational and matter fields, which we
call $\xi$ and $\varphi$ respectively, follow the equation of motion of a free scalar
field with a time dependent mass in a static spacetime of 1+1 dimensions. Consistent
with our restriction to the inhomogeneous sector, we consider both fields already
devoid of zero modes. Now, introducing for these fields creation and annihilation-like
variables defined like one would naturally do in the case of free massless scalar
fields, one reaches a Fock quantization whose evolution is indeed
unitary \cite{ccm1,ccm2} and such that the vacuum is invariant under $S^1$
translations, which is the gauge group of the reduced system. Moreover,
it has been shown that these two natural properties of unitary dynamics and vacuum
invariance in fact pick up this Fock quantization as the unique acceptable one,
up to unitary equivalence
\cite{ccmv1,cmv}. So, taking into account this result for the totally deparameterized
model, we adopt the suitable field parameterization of Refs. \cite{ccm1,ccm2}
both for the gravitational and matter inhomogeneities of our current model,
and describe them in terms of the
creation and annihilation-like variables mentioned above, in order to eventually
carry out the corresponding Fock quantization of the inhomogeneous sector.
We will call these variables $(a^{(\alpha)\ast}_m,a^{(\alpha)}_m)$, with
$m\in\mathbb{Z}-\{0\}$ and $\alpha=\xi,\varphi$.

On the other hand, the homogeneous sector describes Bianchi I cosmologies
with spatial three-torus topology and with a minimally coupled
homogeneous massless scalar field, given by the zero mode of $\Phi$. From now on,
we call $\phi$ this {\sl homogeneous} matter field and $P_{\phi}$ its momentum.
Since the homogeneous sector is to be quantized using LQC methods, we describe the
gravitational variables of this sector in the Ashtekar-Barbero formalism.
Using a diagonal gauge, the nontrivial
components of the densitized triad are $p_{j}/4\pi^{2}$, with $j=\theta, \sigma,
\delta$, whereas those of the $su(2)$ connection are $c_{j}/2{\pi}$ (see e.g. \cite{mbmmp}).
These variables satisfy $\left\{c_{i},p_{j} \right\}=8\pi G \gamma
\delta_{ij}$, where $\gamma$ is the Immirzi parameter and $G$ is the Newton constant
(throughout the text, we set the speed of light equal to the unity).

Two global constraints still remain on this reduced system: the spatial average of
the densitized Hamiltonian constraint, $\mathcal{C}$, and the generator of
$S^1$ translations, $\mathcal{C}_\theta$. On the one hand, $\mathcal{C}$ can be split
into two terms, $\mathcal{C}=\mathcal{C}_{\text{hom}}+ \mathcal{C}_{\text{inh}}$,
the former involving only the homogeneous sector. Then, $\mathcal{C}_{\text{inh}}$
couples the homogeneous gravitational sector with both the gravitational and matter
inhomogeneous sectors through two identical terms, one per field,
$\mathcal{C}_{\text{inh}}=\mathcal{C}_{\text{inh}}^\xi+\mathcal{C}_{\text{inh}}^\varphi$,
where $\mathcal{C}_{\text{inh}}^\xi$ denotes the corresponding coupling term for
the vacuum case~\cite{hybrid2}. The homogeneous term $\mathcal{C}_{\text{hom}}$
is the densitized Hamiltonian constraint of the Bianchi I model minimally coupled to a
homogeneous massless scalar field \cite{mbmmp,awe1}. Thanks to the presence of matter,
the classical Bianchi I model admits solutions of the FRW type. On the other hand,
as it happens to be the case with the inhomogeneous term $\mathcal{C}_{\text{inh}}$,
$\mathcal{C}_\theta$ is the sum of two identical contributions,
$\mathcal{C}_{\theta}=\mathcal{C}_{\theta}^\xi+\mathcal{C}_{\theta}^\varphi$,
where $\mathcal{C}_{\theta}^\xi$ denotes the analog constraint in vacuo \cite{hybrid2}.
We see that, with our choice of variables, matter
and gravitational inhomogeneities contribute in the same way to the constraints,
and then it is straightforward to promote $\mathcal{C}_{\theta}$ and $\mathcal{C}$ to
operators following the hybrid quantization developed for the vacuum case.

\section{Quantum model}
\label{s3}

\subsection{Kinematics and constraint operators}

The quantization of the system starts with the introduction of a kinematical Hilbert space
where the basic variables are represented as operators and where the constraints are imposed
quantum mechanically. For this kinematical Hilbert space, $\mathcal{H}_{\text{kin}}$,
a natural selection is the tensor product
of the kinematical Hilbert space of the gravitational sector
$\mathcal{H}_{\text{kin}}^{\text{grav}}$ times the kinematical Hilbert space of the matter sector
$\mathcal{H}_{\text{kin}}^{\text{matt}}$. Both of these spaces are in turn the tensor product
of two spaces corresponding to homogeneous and inhomogeneous sectors, respectively. Physically,
the nontriviality of the system comes from the couplings introduced at the moment of imposing
the quantum constraints.

For the gravitational sector, then, we carry out the hybrid quantization of
Refs. \cite{hybrid3,hybrid4}, namely $\mathcal{H}_{\text{kin}}^{\text{grav}}$ is
the tensor product of the kinematical Hilbert space of the Bianchi I model in LQC
\cite{awe1,hybrid4}, $\mathcal{H}_{\text{kin}}^{\text{BI}}$, times the standard Fock
space for the inhomogeneities, $\mathcal{F}^{\xi}$, defined in terms of the annihilation
and creation variables previously described for the field $\xi$. The
homogeneous matter sector, on the other hand, is formed by the zero modes of
the massless scalar field and its momentum, determined by $\phi$ and $P_{\phi}$.
In analogy with the nonvacuum cases analyzed in homogeneous LQC (in particular the
Bianchi I model minimally coupled to a massless scalar \cite{awe1}), we take
the standard representation for these variables, choosing $L^{2}(\mathbb{R},d\phi)$
as the Hilbert space.
Finally, since matter and gravitational inhomogeneities have
identical behavior, the kinematical Hilbert space accounting for the matter
inhomogeneities, $\mathcal{F}^{\varphi}$, is totally analogous to $\mathcal{F}^{\xi}$.
Summarizing,
\begin{equation}
\mathcal{H}_{\text{kin}}=\mathcal{H}_{\text{kin}}^{\text{BI}}\otimes
L^{2}(\mathbb{R},d\phi)\otimes\mathcal{F}^{\xi}\otimes\mathcal{F}^{\varphi}.
\end{equation}

For the inhomogeneous sector, the chosen representation is obtained by
promoting the classical variables
$a^{(\alpha)\ast}_m$ and $a^{(\alpha)}_m$, with $m\in\mathbb{Z}-\{0\}$
and $\alpha\in\{\xi,\varphi\}$, to creation and annihilation operators,
$\hat{a}^{(\alpha)\dagger}_{m}$ and $\hat{a}^{(\alpha)}_{m}$,
respectively. With them, it is straightforward to construct the quantum
counterpart of the constraint $\mathcal{C}_{\theta}$,
for which we choose normal ordering. The result is
\cite{hybrid2,hybrid3}
\begin{equation}
\label{eq:dco}
\widehat{\mathcal{C}}_{\theta}=\sum_{m=1}^{\infty}m\widehat{X}^{\xi}_{m}+
\sum_{m=1}^{\infty}m\widehat{X}^{\varphi}_{m}, \quad \qquad
\widehat{X}_{m}^{\alpha}=\hat{a}^{(\alpha)\dagger}_{m}
\hat{a}^{(\alpha)}_{m}-\hat{a}^{(\alpha)\dagger}_{-m}
\hat{a}^{(\alpha)}_{-m}.
\end{equation}

The same strategy is adopted when representing the inhomogeneous
contributions to the coupling terms $\mathcal{C}_{\text{inh}}^\alpha$.
It turns out that the inhomogeneities of the
field $\alpha$ ($\xi$ or $\varphi$) appear in $\mathcal{C}_{\text{inh}}^\alpha$
only via two different quadratic combinations,
$H^{\alpha}_{0}$ and $H^{\alpha}_{\text{int}}$,
whose normal ordered quantum counterparts are
\cite{hybrid2,hybrid3}
\begin{equation}
\widehat{H}_{0}^{\alpha}=\sum_{m=1}^{\infty}m
\widehat{N}^{\alpha}_{m}, \qquad \widehat{H}_{\text{int}}^{\alpha}=
\sum_{m=1}^{\infty}\frac{1}{m}\left(\widehat{N}^{\alpha}_{m}+
\hat{a}^{(\alpha)\dagger}_{m}
\hat{a}^{(\alpha)\dagger}_{-m} + \hat{a}^{(\alpha)}_{m}
\hat{a}^{(\alpha)}_{-m}\right),
\end{equation}
with $\widehat{N}^{\alpha}_{m}=\hat{a}^{(\alpha)\dagger}_{m}
\hat{a}^{(\alpha)}_{m} + \hat{a}^{(\alpha)\dagger}_{-m}
\hat{a}^{(\alpha)}_{-m}$. The above operators $\widehat{X}_{m}^{\alpha}$,
$\widehat{H}^{\alpha}_{0}$, and $\widehat{H}^{\alpha}_{\text{int}}$ act
nontrivially on $\mathcal{F}^{\alpha}$ and have as a common
dense domain the space of $n$-particle states.
We call $n_m^\alpha$ the number of particles of the field $\alpha$ in the mode $m$.

On the other hand, for the homogeneous sector, the basic matter variables are
represented by the operators $\hat{\phi}$, which acts by multiplication, and
$\hat{P}_{\phi}=-i\hbar\partial_\phi$, while
for the gravitational part we adopt the operator
representation discussed in detail in Ref. \cite{hybrid3} (see also Ref.
\cite{hybrid4}), adhering to the
improved dynamics scheme put forward by Ashtekar and Wilson-Ewing \cite{awe1}
(and which was called ``case B'' in Ref. \cite{hybrid3}). Let us briefly
review this quantization scheme. First we recall that, on
$\mathcal{H}_{\text{kin}}^{\text{BI}}$, the operators
$\hat{p}_{i}$ ($i=\theta,\sigma,\delta$), which represent the nontrivial
coefficients of the densitized triad of the Bianchi I model, have a discrete
spectrum equal to the real line. The corresponding eigenstates,
$|p_\theta,p_\sigma,p_\delta\rangle$, form an orthonormal basis (in the
discrete norm) of $\mathcal{H}_{\text{kin}}^{\text{BI}}$. Owing to this
discreteness, there is no well-defined operator representing the connection,
but rather its holonomies. The representation of the matrix elements of these
holonomies incorporates the so-called improved dynamics prescription, which
states that there exists a dynamical (state dependent) minimum length
$\bar\mu_i$ for the straight edges in the $i$th-direction along which the
holonomies are computed. We use the specific improved dynamics
prescription put forward in Ref. \cite{awe1}.
Then, the elementary operators which represent the matrix elements of
the holonomies, called $\hat{\mathcal{N}}_{\bar{\mu}_{i}}$, produce all a constant
shift in the physical Bianchi I volume \cite{awe1,hybrid3}.
The resulting action of
$\hat{\mathcal{N}}_{\bar{\mu}_{i}}$ on the states
$|p_\theta,p_\sigma,p_\delta\rangle$ is quite involved. In order
to simplify the analysis, it is convenient to relabel the basis states
in the form $|v,\lambda_\sigma,\lambda_\delta\rangle$, where $v$ is an
affine parameter proportional to the volume of the compact spatial section,
such that any of the operators $\hat{\mathcal{N}}_{\pm\bar{\mu}_{i}}$
($i=\theta,\sigma,\delta$) causes a unit (positive or negative) shift on it.
The parameters $\lambda_i$ are
all equally defined in terms of the corresponding parameters $p_i$, and verify
that $v=2\lambda_\theta\lambda_\sigma\lambda_\delta$ (see the explicit definitions
in Ref. \cite{awe1}).

Employing the basic homogeneous gravitational operators $\hat{p}_{i}$ and
$\hat{\mathcal{N}}_{\pm\bar{\mu}_{i}}$, we can complete the construction of the
constraint operator $\widehat{\mathcal{C}}=\widehat{\mathcal{C}}_{\text{hom}}+
\widehat{\mathcal{C}}_{\text{inh}}$
exactly in the same way as in the vacuum case \cite{hybrid3}.
This densitized Hamiltonian constraint operator is formed by
\begin{align}
\label{eq:dhco2}
&\widehat{\mathcal{C}}_{\text{hom}}=-
\sum_{i\neq j }\sum_{j }
\frac{\widehat{\Theta}_{i}\widehat{\Theta}_{j}} {16\pi G \gamma^{2}}-
\frac{\hbar^{2}}{2}
\left[\frac{\partial}{\partial \phi}\right]^{2},\\
\label{eq:dhco3}
&\widehat{\mathcal{C}}_{\text{inh}}= 2\pi\hbar\widehat{|p_{\theta}|}
\left(\widehat{H}_{0}^{\xi}+ \widehat{H}_{0}^{\varphi} \right)
+\hbar\widehat{\left[\frac{1}
{|p_{\theta}|^{\frac{1}{4}}}\right]}^{2}
\frac{\left(\widehat{\Theta}_{\delta}+\widehat{\Theta}_{\sigma}
\right)^{2}}{16\pi \gamma^{2}}
\widehat{\left[\frac{1}{|p_{\theta}|^{\frac{1}{4}}}\right]}^{2}
\left(\widehat{H}_{\text{int}}^{\xi}+
\widehat{H}_{\text{int}}^{\varphi} \right),
\end{align}
with $i,j\in\{\theta,\delta,\sigma\}$.
Here, $\widehat{\left[1/{|p_{\theta}|^{{1}/{4}}}\right]}$ is a
regularized triad operator which has a diagonal action on the considered basis of
states. On the other hand,
the operator $\widehat{\Theta}_{i}$ is the quantum counterpart of
the classical quantity $c_ip_i$ and its action on the basis states is
highly nontrivial. In particular, $\widehat{\Theta}_{i}$ and
$\widehat{\Theta}_{j}$ do not commute for $i\neq j$.
We will not give here the explicit action of these operators on our
basis states (which can be found in Ref. \cite{hybrid3}).
Instead, in the following section, we will write down
explicitly the general equation
that must be satisfied by the solutions of the quantum
densitized Hamiltonian constraint.

The above constraint operator leaves
invariant certain subspaces of $\mathcal{H}_{\text{kin}}$, which provide superselection
sectors \cite{abl,hybrid3}. When symmetrizing $\widehat{\mathcal{C}}$, we have chosen a
specific factor ordering which leads to superselection sectors which are particularly
simple and with most convenient properties.
More precisely, instead of considering
$\mathcal{H}_{\text{kin}}^{\text{BI}}$, we can restrict the
homogenous gravitational sector to be the
completion with respect to the discrete norm of the
space spanned by the states $|v,\lambda_\sigma,\lambda_\delta\rangle$
such that $v$, $\lambda_\sigma$, and $\lambda_\delta$ belong to an octant,
for instance $v,\lambda_\sigma,\lambda_\delta >0$
(case on which we will focus our attention from now on), and with $v$ belonging then to
any semilattice $\mathcal{L}_{\epsilon}$ of step four included in $\mathbb{R}^+$:
\begin{equation}
\mathcal{L}_{\epsilon}=\left\{\epsilon + 4k; \ k \in \mathbb{N}  \right\}.
\end{equation}
In this expression, $\epsilon$ is any number in the interval $(0,4]$, and provides the
minimum value that $v$ takes.
In addition, given $\epsilon$,
the labels $\lambda_{a}$ ($a=\sigma$ or $\delta$) are restricted to sectors of the form
$\lambda_{a}=\lambda^{\star}_{a}\omega_{\epsilon}$, where
the $\lambda_{a}^{\star}$'s are any two fixed positive numbers and
$\omega_{\epsilon}$ runs over the following numerable and dense subset of $\mathbb{R}^{+}$:
\begin{equation}
\left\{\left(\frac{\epsilon-2}{\epsilon}\right)^{z}
\prod_{k}\left(\frac{\epsilon+2m_k}{\epsilon+2n_{k}} \right)^{p_{k}}\right\}.
\end{equation}
Here $m_{k},n_{k},p_{k}\in \mathbb{N}$, and $z \in \mathbb{Z}$ when $\epsilon>2$, while $z=0$
otherwise \cite{hybrid3}.

Once we restrict the study to any of the above
superselection sectors, the null eigenspace of the homogeneous densitized triad
operator (which is a proper subspace of $\mathcal{H}_{\text{kin}}^{\text{BI}}$)
ceases to be included in our theory. As a consequence, there is no analog of
the classical cosmological singularity in the quantum model anymore. In this sense,
it is ensured that the singularity is resolved, already at the kinematical level.

\subsection{Physical Hilbert space}

Once we have constructed the constraint operators, we can proceed to determine
the physical states, which must be annihilated by these constraints. Notice that
the two constraint operators commute and can hence be imposed consistently.

Let us consider first, e.g., the $S^1$ symmetry generated by
$\widehat{\mathcal{C}}_\theta$, which amounts to the following
condition
\begin{equation}
\label{eq:mom}
\sum_{m=1}^{\infty}m(X^{\xi}_{m}+ X^{\varphi}_{m})=0, \qquad
X^{\alpha}_{m}=n^{\alpha}_{m}-n^{\alpha}_{-m}, \qquad
\alpha=\xi,\varphi.
\end{equation}
The states that satisfy this condition form a proper subspace of
$\mathcal{F}^\xi\otimes\mathcal{F}^\varphi$, which we call
$\mathcal{F}_{\text{p}}$.

The Hamiltonian constraint operator imposes a more complicated
condition, mainly because of the nontrivial actions of both
$\widehat{\Theta}_i$ on the homogeneous gravitational sector and
$\widehat{H}^\alpha_\text{int}$ on the inhomogeneous sector
\cite{hybrid3,hybrid4}. For our purposes here, it suffices to make
explicit the action of the Hamiltonian constraint operator on
just the homogeneous sector. With this aim, it proves
convenient to introduce an alternate labeling of the basis states of
$\mathcal{H}_{\text{kin}}^{\text{BI}}$. The new labeling is given
by $|v,\Lambda, \Upsilon \rangle$, where
$\Lambda={\ln(\lambda_{\sigma}\lambda_{\delta})}$ and $\Upsilon =
\ln(\lambda_{\delta}/\lambda_{\sigma})$. Next, we expand a
general state $|\Psi\rangle$ in this basis:
\begin{equation}
|\Psi\rangle=\sum_{\bar{v},\bar{\Lambda}, \bar{\Upsilon}}
|\Psi(\bar{v},\bar{\Lambda}, \bar{\Upsilon})\rangle
\otimes|\bar{v},\bar{\Lambda}, \bar{\Upsilon}\rangle.
\end{equation}
Here, $\bar{v}$, $\bar{\Lambda}$, and $\bar{\Upsilon}$ take values
in the corresponding superselection sectors. Let us clarify that
the kets $|\Psi(\bar{v},\bar{\Lambda}, \bar{\Upsilon})\rangle$ are
actually not wave function coefficients, but rather {\sl states}
inasmuch as we have not expanded $|\Psi\rangle$ in a basis of the whole
kinematical Hilbert space, but only of the homogeneous gravitational sector.
On the other hand,
based on our experience with other similar cosmological models,
the {\sl states}
$|\Psi(\bar{v},\bar{\Lambda}, \bar{\Upsilon})\rangle$ for the solutions
of the Hamiltonian constraint are not expected
to be normalizable in
$L^2(\mathbb{R},d\phi)\otimes\mathcal{F}^{\xi}\otimes\mathcal{F}^{\varphi}$,
but rather to belong to a larger space from which one should construct the
physical Hilbert space of the theory. Acting with $\hat{\mathcal{C}}$ on
$|\Psi\rangle$ and projecting over $\langle
v,\Lambda,\Upsilon|$, we obtain:
\begin{align}
\nonumber & -
\frac{8}{\pi  G}
\left[\frac{\partial}{\partial \phi} \right]^{2}
|\Psi\left(v,\Lambda,\Upsilon \right)\rangle +\sum_{\kappa\in\{0,4\}}
\sum_{s\in\{+,-\}}
x_{\kappa}^{s}(v)|\Psi_{\kappa}^{s}
\left(s\kappa+v,\Lambda,\Upsilon \right)\rangle\\
\nonumber& \ -4 \beta e^{2\Lambda} b^2(v)\left[\widehat{H}_{\text{int}}^{\xi}+
\widehat{H}_{\text{int}}^{\varphi} \right]
\sum_{\kappa\in\{0,4\}} \sum_{s\in\{+,-\}}b^2(s\kappa+v)
\frac{s\kappa+v}{v}
x_{\kappa}^{s}({v})
|\Psi_{\kappa}^{s \prime}
\left(s\kappa+v,\Lambda,\Upsilon \right)\rangle \\
& \ +\frac{8}{\beta}\frac{{v}^{2}}{e^{2\Lambda} }
\left[\widehat{H}_{0}^{\xi}+\widehat{H}_{0}^{\varphi}\right]
|\Psi\left(v,\Lambda,\Upsilon \right)\rangle=0
\label{eq:adhc}.
\end{align}
Here, $\beta=[G\hbar/(16 \pi^2 \gamma^2 \Delta)]^{1/3}$, with $\Delta$ denoting the
minimum nonzero eigenvalue allowed for the area in LQG \cite{awe1,hybrid3},
and we have defined
\begin{align}
& b(v)= \left| \sqrt{|v+1|} - \sqrt{|v-1|}\right| ,\\
&
x_{\kappa}^{s}(v)=-\frac{e^{i\pi\kappa/4}}{2}|s2+v|\sqrt{v|s\kappa+v|}
\left\{1+\text{sgn}\left(s\left[2+\frac{\kappa}{2}\right]+v\right)\right\}.
\end{align}
On the other hand, the objects $|\Psi_{\kappa}^{s}
\left(s\kappa+v,\Lambda,\Upsilon \right)\rangle$, are linear
combinations of six contributions in the form
\begin{align}
\nonumber |\Psi_{\kappa}^{s}
\left(s\kappa+v,\Lambda,\Upsilon \right)\rangle
&= \sum_{r\in \{1,-1\}}\Big( |\Psi \left(s\kappa+v,\Lambda + w_{v}(s 2),
\Upsilon +r\, w_{v}(s 2)\right)\rangle \\
\nonumber & +|\Psi \left(s\kappa+v,\Lambda + w_{v}(s\kappa), \Upsilon
+ r\, w_{v}(s\kappa) - 2r\,w_{v}(s 2)  \right)\rangle\\
& + |\Psi \left(s\kappa+v,\Lambda + w_{v}(s\kappa)- w_{v}(s 2),
\Upsilon + r\, w_{v}(s\kappa)- r\,w_{v}(s 2)
\right)\rangle \Big),\label{eq:coef1}
\end{align}
where $w_{v}(s n)=\ln(sn+v)-\ln(v)$. The two last lines of Eq.
\eqref{eq:adhc} correspond to the action produced by $\hat{\mathcal{C}}_{\text{inh}}$,
where we have introduced the notation
\begin{align}
\nonumber |\Psi_{\kappa}^{s\prime}
\left(s\kappa+v,\Lambda,\Upsilon \right)\rangle
& = \sum_{r\in \{1,-1\}}\Big( |\Psi \left(s\kappa+v,\Lambda + w_{v}(s\kappa), \Upsilon
+ r\, w_{v}(s\kappa)  \right)\rangle\\
& + |\Psi \left(s\kappa+v,\Lambda + w_{v}(s\kappa),
\Upsilon + r\, w_{v}(s\kappa)- 2r\,w_{v}(s 2)
\right)\rangle \Big).
\end{align}
Condition \eqref{eq:adhc}, coming from the constraint, is a
difference equation in the variable $v$ and can be seen as an evolution equation in this
variable. In the vacuum model, it has been proven that, formally, a(n infinite but countable)
set of initial data on the section given by the minimum value of $v$,
$v_\text{min}=\epsilon\in(0,4]$, completely determines a solution of the
densitized Hamiltonian constraint \cite{hybrid3,hybrid4}. Since the Hamiltonian
constraint of our model and the one in vacuo have identical structure,
the above result applies also to our case. This property allows
us to identify the physical Hilbert space of the system, that we call
$\mathcal{H}_{\text{phys}}$, as the Hilbert space of these initial data.

The resulting physical Hilbert space, taking into account condition \eqref{eq:mom} as well,
is given by $\mathcal{H}_{\text{phys}}=\mathcal{H}_{\text{phys}}^{\text{BI}}\otimes
L^{2}(\mathbb{R},d\phi) \otimes \mathcal{F}_{\text{p}}$,
where $\mathcal{H}_{\text{phys}}^{\text{BI}}$ is the physical Hilbert space of the Bianchi I
model determined in Ref. \cite{hybrid4}. As discussed in that reference,
the inner product that provides this
Hilbert space structure on the space of initial data is obtained by the requirement that
the complex conjugation relations between a complete set of classical observables turn
into adjoint relations between the corresponding operators.

\subsection{{\sl Projection} to LRS-Gowdy}
\label{s3c}

The Gowdy $T^{3}$ model with linear polarization is symmetric under the interchange
of the directions coordinatized by $\sigma$ and $\delta$. Owing to this, it
has a subset of classical solutions with local rotational symmetry (LRS),
in which the scale factors of these two directions can be identified during the entire
evolution. We can then restrict the Gowdy model, both in vacuo and with matter,
to the LRS-Gowdy model in which every solution is of this kind. The restriction can be
performed classically, prior to quantization, or starting with the quantized model.
We will focus our attention on the latter approach, passing from quantum Gowdy
to quantum LRS-Gowdy, and leave for the interested reader the proof that
the quantum model obtained in this way is indeed recovered by a direct quantization
of the classical LRS-Gowdy spacetimes along the lines explained in this work.

In analogy with the discussion of Ref. \cite{awe1}, in which the quantum FRW model is
obtained from quantum Bianchi I, we define the following map from
(generalized) states associated with the Gowdy model to those of the
LRS-Gowdy cosmologies,
denoted by $|\psi(v,\Lambda)\rangle$:
\begin{equation}
\label{eq:map} |\Psi(v, \Lambda, \Upsilon)\rangle \qquad \longrightarrow
\qquad \sum_{\Upsilon}|\Psi(v,\Lambda, \Upsilon)\rangle \equiv|\psi(v,\Lambda)\rangle.
\end{equation}
The sum is carried out over all values of $\Upsilon$ in the considered superselection
sector. Applying this map in the Hamiltonian constraint \eqref{eq:adhc}, we
obtain
\begin{align}\
\nonumber & -
\frac{4}{\pi  G}
\left[\frac{\partial}{\partial \phi} \right]^{2}
|\psi\left(v,\Lambda \right)\rangle
+\sum_{\kappa\in\{0,4\}}
\sum_{s\in\{+,-\}}
x_{\kappa}^{s}(v)|\psi_{\kappa}^{s}
\left(s\kappa+v,\Lambda\right)\rangle\\
\nonumber & \ - 8\beta b^{2}({v})e^{2\Lambda}\left[\widehat{H}_{\text{int}}^{\xi}+
\widehat{H}_{\text{int}}^{\varphi} \right]
\sum_{\kappa\in\{0,4\}} \sum_{s\in\{+,-\}}b^2(s\kappa+v)
\frac{s\kappa+v}{v}
x_{\kappa}^{s}({v})
|\psi\left(s\kappa+v,\Lambda+ w_{v}(s\kappa)\right)\rangle\\
& \ +\frac{4}{\beta}\frac{{v}^{2}}{e^{2\Lambda} }\left[\widehat{H}_{0}^{\xi}+
\widehat{H}_{0}^{\varphi}\right]|\psi\left(v,\Lambda\right)\rangle=0,
\label{eq:adhcLRS}
\end{align}
where $|\psi_{\kappa}^{s}(s\kappa+v, \Lambda)\rangle$ are the combinations
\begin{align}
\nonumber |\psi_{\kappa}^{s}\left(s\kappa+v,\Lambda\right)\rangle
&= |\psi \left(s\kappa+v,\Lambda + w_{v}(s\kappa)- w_{v}(s 2)\right)\rangle\\
& +|\psi \left(s\kappa+v,\Lambda + w_{v}(s\kappa)\right)\rangle +|\psi \left(s\kappa+v,
\Lambda + w_{v}(s 2)\right)\rangle.
\label{eq:coeff3}
\end{align}
As we have already remarked, the result agrees with the constraint obtained by a
suitable hybrid quantization of the classical LRS-Gowdy model.
It is worth noting that the introduced map works because the coefficients appearing in
Eq. \eqref{eq:adhc} do not depend on the variable $\Upsilon$, over which one sums to
perform the {\sl projection}. Indeed, this kind of map
only makes sense if the classical model admits the imposition of an additional
symmetry which allows its
reduction into a dynamically stable submodel. A similar {\sl projection}
summing over $\Lambda$ is not viable because
the coefficients of the inhomogeneous contributions in the constraint
depend explicitly on this variable, reflecting the fact that the associated kind
of isotropic solutions exist just when the inhomogeneities are unplugged.

\section{Discussion}
\label{s4}

We have completely quantized the Gowdy $T^{3}$ model with linearly
polarized gravitational waves provided with a minimally coupled massless scalar
field as matter content. The description adopted for the matter inhomogeneities
is such that they can be treated in exactly the same way as the gravitational
ones, the former just duplicating the contributions of the latter in the constraints.
In this situation, we have been able to apply the hybrid quantization methods
developed in Refs. \cite{hybrid1,hybrid2,hybrid3,hybrid4} almost straightforwardly to
this system with local physical degrees of freedom both in the matter content and in
the gravitational field. To our knowledge, it is the first time that a model with
these properties has been quantized to completion in the framework of LQC.

Since the structure of the constraints when the matter field is present
is the same as in vacuo, all the results obtained in Ref.
\cite{hybrid3} for the vacuum Gowdy model apply as well to our model.
Thus, in particular, we recover on physical states the standard quantum
field theory description of both the matter and the gravitational
inhomogeneities, living on a cosmological background quantized using LQC methods
and consisting of a Bianchi I universe with a {\sl homogeneous}
massless scalar field.
In addition, it is guaranteed that the states which are the analog of
the classical singularity decouple naturally in the quantum model,
so that, to this extent, the initial singularity is resolved at the kinematical
level.

Conceptually, the hybrid quantization of the present family of inhomogeneous cosmologies
has introduced no technical complication with respect to the vacuum case.
Nonetheless, the situation is radically different when one considers the interest of the
quantum model from a physical point of view. In fact, thanks to the inclusion of the
massless scalar field, the homogeneous sector of the Gowdy $T^3$ model, namely the
Bianchi I model, admits now isotropic flat solutions of FRW type, while in vacuo only
the trivial Minkowskian solution is allowed.

On the other hand, the analysis of the
classical solutions of the linearly polarized Gowdy $T^{3}$
model in vacuo \cite{m} and the study of the effective dynamics obtained from the
hybrid quantization of this model \cite{gow-eff}
show that small inhomogeneities do not increase arbitrarily in
the evolution. Then, if we consider initial data which
are sufficiently close to homogeneity, the
corresponding solution would remain approximately homogeneous during the evolution.
Besides, in the nonvacuum
model, matter and gravitational inhomogeneities evolve in identical way. This strongly
indicates that initial data in a sufficiently small neighborhood of those with isotropy and
homogeneity have to lead to approximately isotropic and homogeneous solutions.
Therefore, it is natural to compare the dynamics of our Gowdy model with that of the
flat FRW model (with three-torus topology) in order to see how the inclusion of
anisotropies and inhomogeneities affects the evolution of a flat FRW background.
Moreover, we are now in a perfect situation to carry out this comparison at
the quantum level, since here we have accomplished the full quantization of the
Gowdy $T^3$ model in the presence of the massless matter, and the loop
quantization of the FRW model coupled to the homogeneous massless field is
well known \cite{aps,mmo}. Even though the inhomogeneities in our model
are not all those allowed in a universe like the one which we observe
(but just a subfamily with the symmetries of the Gowdy $T^3$ cosmologies),
their analysis should shed light on the kind of quantum effects affecting
the evolution and on the consequences of the quantum geometry on
the primordial fluctuations.

For these purposes, it is preferable to focus on the LRS-Gowdy model
derived in Sec. \ref{s3c}. Indeed, the consideration of the two degrees
of freedom of anisotropy that the homogeneous sector of the general Gowdy model
possesses would only complicate the equations unnecessarily. The presence of
either two degrees or just one degree of anisotropy does not seem to
have any conceptual relevance for the proposed analysis.

In order to face this analysis, the idea is to add and subtract
in Eq. \eqref{eq:adhcLRS} the term corresponding to the FRW model, which coincides
with the first line of Eq. \eqref{eq:adhcLRS} but keeping the variable $\Lambda$
unchanged. We can then rewrite Eq. \eqref{eq:adhcLRS} as the constraint equation of
the FRW model coupled to a homogeneous massless scalar field plus a number of
contributions coming from all other terms. These contributions contain the
inhomogeneities and the difference between the gravitational parts of the
constraints for LRS-Bianchi I and FRW, a difference which is due to the anisotropies.
In this way, the resulting expression modifies
the densitized Hamiltonian constraint of the FRW model
by the effects of the anisotropies and inhomogeneities, so that it is not longer
equal to zero. As we have commented, we are interested in comparing the FRW model
with the inhomogeneous LRS-Gowdy model when these inhomogeneities and anisotropies are small.
In this regime, it makes sense to apply a type of Born-Oppenheimer approximation, similar
to others commonly employed in cosmology (see e.g. \cite{hh,rv}), and
assume that the variations
of the isotropic degrees of freedom and those of the rest of degrees have considerably
different typical scales, therefore giving them a different status. Then, in this
approximation, it is easier to derive effectively the influence that anisotropies and
inhomogeneities produce on the isotropic background. We leave for future research
this detailed analysis.

\section*{Acknowledgements}

The authors are grateful to L. Garay and J. Olmedo for discussions. This work was
supported by the Spanish MICINN Project No. FIS2008-06078-C03-03 and
the Consolider-Ingenio Program CPAN No. CSD2007-00042. M. M-B and D. M-dB have been
supported by CSIC and the European Social Fund under the grants I3P-BPD2006 and JAEPre\_09\_01796,
respectively.

\end{document}